\documentclass{article}
\usepackage{amsmath,amsfonts,amssymb,setspace,cancel,url,hyperref,verbatim,color,todonotes}
\usepackage[a4paper]{geometry}
\usepackage{breqn,caption}
\usepackage{cite}
\newcommand{\be}{\begin{equation}}
\newcommand{\ee}{\end{equation}}

\newcommand{\ON}[1]{\mathrm{O}( #1 )}
\newcommand{\SU}[1]{\mathrm{SU}( #1 )}
\newcommand{\SL}[1]{\mathrm{SL}( #1 )}
\newcommand{\GL}[1]{\mathrm{GL}( #1 )}
\newcommand{\SO}[1]{\mathrm{SO}( #1 )}
\newcommand{\Spin}[1]{\mathrm{Spin}(#1)}

\newcommand{\USp}[1]{\mathrm{USp}(#1)}

\newcommand{\TA}{{\cal A}}
\newcommand{\TB}{{\cal B}}
\newcommand{\TC}{{\cal C}}
\newcommand{\TD}{{\cal D}}

\newcommand{\Fa}{\mathcal{F}}
\newcommand{\Fb}{\mathcal{H}}
\newcommand{\Fc}{\mathcal{J}}

\newcommand{\D}{\mathfrak{D}}

\newcommand{\bmu}{\bar{\mu}}

\newcommand{\bi}{\bar{i}}

\newcommand{\tomega}{\tilde{\omega}}

\newcommand{\tnabla}{\tilde{\nabla}}

\newcommand{\mbf}[1]{\mathbf{#1}}
\newcommand{\gL}{\mathcal{L}}

\newcommand{\gM}{\mathcal{M}}
\newcommand{\gH}{\mathcal{H}}
\newcommand{\cR}{\mathcal{R}}

\newcommand{\tV}{\tilde{V}}
\newcommand{\bae}{\bar{e}}
\newcommand{\bag}{\bar{g}}
\numberwithin{equation}{section}

\begin{document}

\begin{titlepage}
\vfill

\begin{flushright}
LMU-ASC 62/16
\end{flushright}

\vfill

\begin{center}
   \baselineskip=16pt
   	{\Large \bf From Exceptional Field Theory to Heterotic Double Field Theory via K3}
   	\vskip 2cm
 	{\large \bf Emanuel Malek} 
   	\vskip .6cm
   	{\it Arnold Sommerfeld Center for Theoretical Physics, Department f\"ur Physik, \\ Ludwig-Maximilians-Universit\"at M\"unchen, Theresienstra{\ss}e 37, 80333 M\"unchen, Germany}
   	\vskip .6cm
   	{{E.Malek@lmu.de}} \\
   	\vskip 2cm
\end{center}
\vfill

\begin{abstract}
In this paper we show how to obtain heterotic double field theory from exceptional field theory by breaking half of the supersymmetry. We focus on the $\SL{5}$ exceptional field theory and show that when the extended space contains a generalised $\SU{2}$-structure manifold one can define a reduction to obtain the heterotic $\SO{3,n}$ double field theory. In this picture, the reduction on the $\SU{2}$-structure breaks half of the supersymmetry of the exceptional field theory and the gauge group of the heterotic double field theory is given by the embedding tensor of the reduction used. Finally, we study the example of a consistent truncation of M-theory on K3 and recover the duality with the heterotic string on $T^3$. This suggests that the extended space can be made sense of even in the case of non-toroidal compactifications.
\end{abstract}

\vfill

\setcounter{footnote}{0}
\end{titlepage}

\tableofcontents

\newpage

\section{Introduction} \label{s:Introduction}
Exceptional field theory \cite{Hull:2007zu,Berman:2010is,Hohm:2013uia} is an $E_{d(d)}$-manifest extension of supergravity which has been shown to include 11-dimensional and IIB SUGRA in one unified formalism. The starting point for exceptional field theory (EFT), just as for generalised geometry \cite{Coimbra:2011nw,Coimbra:2012af}, is a Kaluza-Klein split of 11-dimensional SUGRA. The bosonic and fermionic degrees of freedom then form representations of the exceptional groups and their maximal compact subgroups, respectively. In EFT, one also extends the coordinates to form a representation of the exceptional groups.

When one considers the $E_{d(d)}$ EFT on a $d$-torus, the interpretation of the exceptional group and the extra coordinates becomes very clear. The exceptional group, or rather its integer part $E_{d(d)}(\mathbb{Z})$, represents the U-duality group, and the extended coordinates are the Fourier duals of momentum- and wrapping-modes of branes. However, on more general backgrounds the interpretation of the extended coordinate space is much less clear. However, in the case of double field theory (DFT) \cite{Hull:2009mi,Hull:2009zb}, which is an $\ON{d,d}$-manifest extension of type II SUGRA following in the footsteps of earlier work \cite{Duff:1989tf,Duff:1990hn,Siegel:1993th,Siegel:1993xq,Tseytlin:1990nb,Tseytlin:1991wr}, one can interpret the doubled space as arising from the independent zero-modes of the left- and right-movers of the string, and one might expect a similar picture in the case of EFT.

With this in mind, we here wish to study exceptional field theories on backgrounds with non-trivial structure group. To be concrete we work with the $\SL{5}$ EFT which has a seven-dimensional ``external'' space, and a 10-dimensional ``extended internal'' space. In this case one can consider a background which has generalised $\SU{2}$-structure \cite{Malek:2016bpu,Coimbra:2014uxa} in which case the background breaks half of the supersymmetry of the exceptional field theory. A particular example of such a background would be a K3 surface.

In \cite{Malek:2016bpu} the distinction is drawn between the linear symmetry group of a theory and the duality group of the truncation on a particular background. The linear symmetry group of a theory determines what representations its field content transforms under. For example the linear symmetry group is $\GL{d}$ in the case of $d$-dimensional general relativity, or $E_{d(d)}$ in the case of exceptional generalised geometry and EFT, because of the inclusion of $p$-form field strengths.

On the other hand, the duality group should here be understood as the symmetry group acting on the moduli space of a truncation of the theory on a particular background. It is this group which becomes the global symmetry group of the lower-dimensional gauged SUGRA and in principle this is different from $E_{d(d)}$, and even much larger. In particular, for seven-dimensional consistent ${\cal N}=2$ truncations of EFT, \cite{Malek:2016bpu} shows that the duality group becomes $\ON{3,n}$ where $n \neq 3$ in general. When the background is generalised parallelisable, the linear symmetry and duality groups coincide explaining the emergence of the $E_{d(d)}$ groups as the global symmetry group for maximally supersymmetric consistent truncations. We will show that the extended coordinate space can be understood to enhance in a similar fashion with the duality group.

In particular, we use the technology of \cite{Malek:2016bpu} to show that exceptional field theory can be reduced to the heterotic double field theory \cite{Siegel:1993th,Siegel:1993xq,Hohm:2011ex} when the extended space contains a generalised $\SU{2}$-structure manifold. The generalised $\SU{2}$-structure breaks half of the supersymmetry and the embedding tensor of the particular $\SU{2}$-structure reduction that is used defines the gauging of the heterotic double field theory. This is reminiscent of the procedure used in \cite{Ciceri:2016dmd} to reduce exceptional field theory to massive IIA supergravity.

From our work a picture emerges for the role of the extended coordinates of EFT on such $\SU{2}$-structure manifolds. A twisted version of the extended coordinates can be used to define the $\left(n+3\right)$-dimensional extended space of the $\ON{3,n}$ heterotic double field theory. Furthermore, the generalised Lie derivative of the $\SL{5}$ EFT reduces to the $\ON{3,n}$ heterotic DFT, with the gauging determined by the reduction on the $\SU{2}$-structure space. Indeed, the entire action reduces to that of the heterotic $\ON{3,n}$ DFT in the so-called ``frame formalism'' \cite{Hohm:2010xe,Hohm:2011ex}.

We also use these results to show how the duality between M-theory on K3 and the heterotic string on $T^3$ emerges in exceptional field theory. The K3 surface depends on four coordinates, which when chosen as part of the section, become the M-theory coordinates. On the other hand, if the section is chosen to exclude these four coordinates, we obtain a truncation of the abelian $\ON{3,19}$ heterotic DFT on $T^3$. Thus, a duality here corresponds to a change of section, in the same way that a conventional U-duality does, as has been advocated in \cite{Berman:2014hna,Berman:2014jsa}. In the language of \cite{Hull:2006va}, this corresponds to a choice of polarisation.

We begin with a short review of the $\SL{5}$ EFT in section \ref{s:Overview} and a summary of the relevant findings of \cite{Malek:2016bpu} in section \ref{s:N2Summary}. In section \ref{s:HetDeformation} we then show how to perform the reduction Ansatz that gives rise to the heterotic DFT. We also discuss how $\left(n+3\right)$-dimensional extended space emerges and how the $\SL{5}$ generalised Lie derivative reduces to the heterotic one. In section \ref{s:HetAction} we show that the $\SL{5}$ EFT action reduces to that of the heterotic DFT in the frame formalism \cite{Hohm:2010xe} and with a Kaluza-Klein split \cite{Hohm:2013nja} with the seven external dimensions. Finally, we discuss in section \ref{s:MHetDuality} how the duality between M-theory on K3 and the heterotic string on $T^3$ arises in exceptional field theory before concluding in section \ref{s:Conclusions}.

\section{Overview of exceptional field theory} \label{s:Overview}
Here we will give a very brief overview of the main ingredients of the $\SL{5}$ exceptional field theory which we will require in the remaining discussion. We will introduce further concepts where they are needed along the way in the rest of the paper. We refer the reader to the reviews \cite{Berman:2013eva,Aldazabal:2013sca,Hohm:2013bwa} and the papers \cite{Berman:2010is,Berman:2011cg,Hohm:2013uia} for more details.

The $\SL{5}$ EFT can be viewed as a reformulation of 11-dimensional supergravity which makes the linear symmetry group $\SL{5}$ manifest. Thus, the starting point is 11-dimensional supergravity in a 7+4 split. Let us use $x^\mu$, $\mu = 1, \ldots, 7$, as coordinates for the ``external'' 7-d space and label $y^{\bar{i}}$, $\bi = 1, \ldots, 4$ as the four ``internal coordinates''. These are part of 10 ``extended coordinates'', $Y^{ab}$, forming the antisymmetric representation of $\SL{5}$, where we use $a, b = 1, \ldots, 5$ as fundamental $\SL{5}$ indices. In the case where the internal geometry really is a torus, the extra six coordinates can be understood as being dual to wrapping modes of branes.

All scalars with respect to this $\left(7+4\right)$-split can be described by the generalised metric
\begin{equation}
 \gM_{ab} \in \SL{5} / \USp{4}\,.
\end{equation}
Similarly all bosonic objects with one leg in the external space can be combined into 10 vector fields $\TA_\mu{}^{ab}$, those with two external legs can be combined into five two-forms $\TB_{\mu\nu,a}$, etc.

The local symmetries of 11-dimensional supergravity, i.e. diffeomorphisms and $p$-form transformations, also combine into a $\SL{5}$ action, generated by the so-called generalised Lie derivative. For a tensor in the $\SL{5}$ fundamental representation $V^a$ of weight $\lambda$ this takes the form \cite{Berman:2011cg,Coimbra:2011ky,Berman:2012vc}
\begin{equation}
 \gL_\Lambda V^a = \frac12 \Lambda^{bc} \partial_{bc} V^a - V^b \partial_{bc} \Lambda^{ac} + \frac15 V^a \partial_{bc} \Lambda^{bc} +\frac{\lambda}{2} V^a \partial_{bc} \Lambda^{bc} \,. \label{eq:LieDerivative}
\end{equation}

For consistency the algebra of generalised diffeomorphisms must close, i.e.
\begin{equation}
 \left[ \gL_{\Lambda_1}, \gL_{\Lambda_2} \right] V^a = \gL_{\left[\Lambda_1,\Lambda_2\right]_D} V^a \,. \label{eq:AlgebraClosure}
\end{equation}
Here the $D$-bracket just represents the action of a generalised Lie derivative,
\begin{equation}
 \left[ \Lambda_1, \Lambda_2 \right]_D^{ab} = \gL_{\Lambda_1} \Lambda_2^{ab} \,.
\end{equation}
In order for \eqref{eq:AlgebraClosure} to hold one needs to impose the so-called section condition \cite{Berman:2011cg,Coimbra:2011ky}
\begin{equation}
 \partial_{[ab} f \partial_{cd]} g = 0 \,, \qquad \partial_{[ab} \partial_{cd]} f = 0 \,,
\end{equation}
where $f$ and $g$ denote any two objects of the $\SL{5}$ EFT. There are two inequivalent solutions to the section condition, one corresponding to 11-dimensional SUGRA and the other corresponds to type IIB. Upon using a solution of the section condition, the generalised Lie derivative \eqref{eq:LieDerivative} generates the $p$-form gauge transformation and diffeomorphisms of the corresponding SUGRA. Similarly, the EFT action reduces to that of 11-dimensional or IIB SUGRA \cite{Blair:2013gqa,Hohm:2013vpa,Hohm:2013uia}, upon imposing a solution of the section condition.

In exceptional field theory, and also double field theory, one can then interpret the choice of section as a duality transformation. In particular, the duality between strings and waves \cite{Berman:2014hna}, branes and monopoles \cite{Berman:2014jsa}, and their non-geometric counterparts \cite{Bakhmatov:2016kfn}, can be seen to arise this way. It has also been suggested that the M-theory / F-theory duality can be interpreted this way \cite{Berman:2015rcc}. Our work here also suggests that the heterotic / M-theory duality can also be seen as an exchange of solutions of the section condition.

\section{Summary of consistent ${\cal N}=2$ truncations} \label{s:N2Summary}

\subsection{Reformulation of the $\SL{5}$ EFT} \label{s:N2Reformulation}
In \cite{Malek:2016bpu} it was shown how to construct seven-dimensional half-maximal consistent truncations of the $\SL{5}$ EFT. We will use this technology here to perform a half-maximally supersymmetric reduction of $\SL{5}$ EFT which yields the $\ON{n,3}$ heterotic double field theory. Let us begin by reviewing the relevant results of \cite{Malek:2016bpu}.

A generalised $\SU{2}$-structure is defined by the nowhere vanishing $\SL{5}$ tensors
\begin{equation}
 \left( \kappa, \quad A_a, \quad A^a, \quad B_{u,ab} \right) \,,
\end{equation}
where $a = 1, \ldots, 5$ is a fundamental $\SL{5}$ index, $u =1, \ldots, 3$ is a triplet index of $\SU{2}_R$ and $\kappa$ is a tensor density under generalised diffeomorphisms, which can be identified with the determinant of the external metric $\kappa = |e|^{1/7}$. These tensors are subject to the compatibility conditions
\begin{equation}
 A^a A_a = \frac12 \,, \qquad B_{u,ab} A^b = 0 \,, \qquad B_{u,ab} B_{v,cd} \epsilon^{abcde} = 4 \sqrt{2} A^e\, \delta_{uv} \,. \label{eq:CompatibilityConditions}
\end{equation}
In \cite{Malek:2016bpu} it was shown that any set of such tensors imply the existence of two nowhere-vanishing spinors and hence a truncation on such a background gives a half-maximally supersymmetric theory. Because these tensors define a $\SU{2}$-structure group and $\SU{2} \subset \USp{4}$ they also implicitly define a generalised metric.

Furthermore, in \cite{Malek:2016bpu}, it was shown how to rewrite the entire EFT in terms of the $\SU{2}$-structure $\left( \kappa,\, A_a,\, A^a,\, B_{u,ab}\right)$ instead of the generalised metric $\gM_{ab}$. These can be thought of as the exceptional analogue of the (almost) K\"ahler and (almost) complex structure of ordinary $\SU{2}$-structure manifolds. To rewrite the action one introduces a generalised $\SU{2}$-connection whose intrinsic torsion can be used to rewrite the generalised Ricci scalar of EFT, as well as the SUSY variations. The intrinsic torsion transforms in the following representations of $\SU{2}_S \times \SU{2}_R \subset \SL{5}$:
\begin{equation}
 W_{int} = 2 \cdot \left(\mbf{1},\mbf{1}\right) \oplus \left(\mbf{3},\mbf{1}\right) \oplus 2 \cdot \left(\mbf{1},\mbf{3}\right) \oplus \left(\mbf{3},\mbf{3}\right) \oplus 3 \cdot \left(\mbf{2},\mbf{2}\right) \oplus \left(\mbf{2},\mbf{4}\right) \,. \label{eq:IntrinsicTorsion}
\end{equation}
We refer readers who are interested in the definition of intrinsic torsion to \cite{Joyce,Coimbra:2014uxa} as well as for this particular case \cite{Malek:2016bpu}.

In order to write the intrinsic torsion explicitly, we will make use of
\begin{equation}
 V_u{}^{ab} = \epsilon^{abcde} B_{u,cd} A_e \,, \qquad \textrm{ and } \qquad \tilde{V}_u{}^{ab} = \kappa \epsilon^{abcde} B_{u,cd} A_e \,,
\end{equation}
where $\tilde{V}_u{}^{ab}$ has weight $\frac15$ and is thus a generalised vector. In addition we will need projectors onto the $\left(\mbf{1}, \mbf{1}\right) \subset \mbf{5}$, $\left(\mbf{3},\mbf{1}\right) \subset \mbf{10}$ as $\SL{5} \rightarrow \SU{2}_S \times \SU{2}_R$ as well as a projector onto the $\left(\mbf{2},\mbf{4}\right) \subset \left(\mbf{2},\mbf{2}\right) \times \left(\mbf{1},\mbf{3}\right)$ of $\SU{2}_S \times \SU{2}_R$. These are given by
\begin{equation}
 \begin{split}
  P_a{}^b &= \delta_a{}^b - 2  A_a A^b \,, \\
  P_{ab}{}^{cd} &= \left( \delta_{ab}{}^{cd} - \frac1{2\sqrt{2}} B_{u,ab} V_u{}^{cd} + 4 A_{[a} A^{[c} \delta_{b]}{}^{d]} \right) \,, \\
  P_{a}{}^{u,b}{}_{v} &= \delta_a{}^b \delta^u{}_v + \frac{\sqrt{2}}{3} B^u{}_{ac} V_v{}^{cb} \,.
 \end{split}
\end{equation}

We can now write down the irreducible components of the intrinsic torsion.
\paragraph{Singlets}
\begin{equation}
 \begin{split}
  S &=  A^a \partial_{ab} A^b \,, \\
  T &= \frac1{12\kappa} \epsilon_{uvw} V^{u,cd} \gL_{\tV^v} B^w{}_{cd} \,.
 \end{split} \label{eq:TSinglets}
\end{equation}

\paragraph{$\left(\mbf{1},\mbf{3}\right)$}
\begin{equation}
 \begin{split}
  T_u &= - 2 \kappa^2 A^a \gL_{\tV_u} \left(A_a \kappa^{-3} \right) \,, \\
  S_u &= 2 \kappa^{-6} \gL_{\tV_u} \kappa^5 \,.
 \end{split} \label{eq:T13}
\end{equation}

\paragraph{$\left(\mbf{3},\mbf{1}\right)$}
\begin{equation}
 \begin{split}
  T_{ab} &= \frac1{12\kappa} P_{ab}{}^{cd} \gL_{\tV_u} B^u{}_{cd} \\
  &= \frac1{12\kappa} \left( \gL_{\tV_u} B^u{}_{ab} - \frac{1}{2\sqrt{2}} B^v{}_{ab} V_v{}^{cd} \gL_{\tV_u} B^u{}_{cd} + 4 A^c A_{[a} \gL_{\tV_u} B^u{}_{b]c} \right) \,. \label{eq:T31}
 \end{split}
\end{equation}

\paragraph{$\left(\mbf{3},\mbf{3}\right)$}
\begin{equation}
 \begin{split}
  T^u{}_{ab} &= \frac{1}{12\kappa} \epsilon^{uvw} P_{ab}{}^{cd} \gL_{\tV_v} B_{w,cd} \\
  &= \frac1{12\kappa} \epsilon^{uvw} \left( \gL_{\tV_v} B_{w,ab} - \frac1{2\sqrt{2}} B^x{}_{ab} V_x{}^{cd} \gL_{\tV_v} B_{w,cd} + 4 A^c A_{[a} \gL_{\tV_v} B_{|w|,b]c} \right) \,. \label{eq:T33}
 \end{split}
\end{equation}

\paragraph{$\left(\mbf{2},\mbf{2}\right)$}
\begin{equation}
 \begin{split}
  S_a &= \frac{1}{\kappa^3} \partial_{ab} \left( A^b \kappa^3 \right) - 2 A_a A^b \partial_{bc} A^c \,, \\
  T_a &= \frac{1}{12\kappa} \epsilon^{uvw} B_{u,ab} V_v{}^{bc} \gL_{\tV_w} A_c \,, \\
  U_a &= \frac1{\kappa} B_{u,ab} \gL_{\tV^u} A^b \,.
 \end{split} \label{eq:TDoublets}
\end{equation}

\paragraph{$\left(\mbf{2},\mbf{4}\right)$}
\begin{equation}
 T^u{}_a = \frac{1}{\kappa} P_{a}{}^{u,b}{}_{v} \epsilon^{vwx} B_{w,bc} \gL_{\tV_x} A^c \,. \label{eq:TDoublets4}
\end{equation}

The generalised Ricci scalar $\cR$ of the EFT is then given by
\begin{equation}
 \begin{split}
  \cR &= 8\, S^2 - 2\, T^2 - 8\sqrt{2}\, ST - 3\, T_u T^u + T_u S^u - \frac34\, S_u S^u - 16\sqrt{2}\, \epsilon^{abcde} T_{ab} T_{cd} A_e \\
  & \quad - 36\sqrt{2}\, \epsilon^{abcde} T^u{}_{ab} T_{u,cd} A_e - \frac{4\sqrt{2}}{3}\, M^{ab} S_a S_b - \frac{16}{3}\, M^{ab} S_a T_b + \frac83\, M^{ab} U_a S_b \,, \label{eq:RicciScalar}
 \end{split}
\end{equation}
which is to be thought of as the half-maximal analogue to the flux formulation of DFT and EFT \cite{Aldazabal:2011nj,Geissbuhler:2011mx,Grana:2012rr,Aldazabal:2013mya,Geissbuhler:2013uka,Berman:2013uda,Blair:2014zba}.
The EFT potential, which is defined as all the terms in the EFT action with only derivatives along $Y^{ab}$, is in turn given by
\begin{equation}
 V = -\frac14 \cR + \frac12 V_u{}^{ab} V^{u,cd} \tnabla_{ab} g_{\mu\nu} \tnabla_{cd} g^{\mu\nu} \,,
\end{equation}
where $\tnabla_{ab}$ is the $\SU{2}$-connection. Because $g_{\mu\nu}$ is a $\SL{5}$ density of weight $\frac25$, the $\SU{2}$-connection acts as
\begin{equation}
 \tnabla_{ab} g_{\mu\nu} = |e|^{2/7} \partial_{ab} \left( g_{\mu\nu} |e|^{-2/7} \right) \,. \label{eq:Nablag}
\end{equation}

Finally, the kinetic terms of the EFT action can also be written in terms of $A^a$, $A_a$ and $B_{u,ab}$ instead of the generalise metric $\gM_{ab}$. In \cite{Malek:2016bpu} it was shown that they are given by
\begin{equation}
 \begin{split}
  L_{kin} &= \frac{1}{2\sqrt{2}} g^{\mu\nu} \left( D_\mu B_{u,ab} D_\nu B^{u}{}_{cd} \right) \epsilon^{abcde} A_e - 14\, g^{\mu\nu} D_\mu A^a D_\nu A_a \\
  & \quad + \frac18 \Fa_{\mu\nu}{}^{ab} \Fa^{\mu\nu\,cd} \left( B_{u,ab} B^u{}_{cd} - B_{u[ab} B^u{}_{cd]} \right) - \frac{1}{48} \Fb_{\mu\nu\rho,a} \Fb^{\mu\nu\rho}{}_b A^a A^b \,,
 \end{split} \label{eq:KinTerms}
\end{equation}
up to terms which vanish in a ${\cal N}=2$ theory.

The full EFT action is then given by
\begin{equation}
 S = \int d^{10}Y d^7x |e| \left( L_{EH} + L_{kin} - V \right) + S_{top} \,, \label{eq:action}
\end{equation}
where $L_{EH}$ is the 7-dimensional external Einstein-Hilbert term and $S_{top}$ is the topological term of the EFT action \cite{Hohm:2013uia,Musaev:2015ces,Bosque:2016fpi} which require no further modification.

\subsection{Consistent ${\cal N}=2$ truncations} \label{s:ConsTruncation}
As argued in \cite{Malek:2016bpu}, the truncation must not keep any doublets of the $\SU{2}$-structure group in order to yield an honest seven-dimensional ${\cal N}=2$ theory. In particular, a nowhere-vanishing doublet of $\SU{2}_S$ would imply that the structure group is actually trivial and there is underlying ${\cal N}=4$ supersymmetry. By decomposing $\SL{5} \rightarrow \SU{2}_S \times \SU{2}_R$ one finds that after removing the $\SU{2}_S$ doublets all remaining fields organise themselves into triplets and singlets of $\SU{2}_S$ and $\SU{2}_R$.

As a result, one can define a ${\cal N}=2$ truncation of the $\SL{5}$ theory by expanding all fields in terms of a basis of sections of the $\left(\mbf{1},\mbf{1}\right)$-, $\left(\mbf{3},\mbf{1}\right)$- and $\left(\mbf{1},\mbf{3}\right)$-bundles of $\SU{2}_S \times \SU{2}_R$. Because $\SU{2}_R$ is trivially fibred there can be only three sections of $\SU{2}_R$ while $\SU{2}_S$ is non-trivially fibred and hence one can use $n$ sections of the $\left(\mbf{3},\mbf{1}\right)$. $n$ determines the number of vector multiplets in the seven-dimensional half-maximal gauged SUGRA. We label these sections as
\begin{equation}
 \rho(Y) \,, \qquad n^a(Y) \,, \quad n_a(Y) \,, \quad \omega_{M,ab}(Y) \,,
\end{equation}
where we have also introduced a $\SL{5}$ density $\rho(Y)$. Here $M = 1, \ldots, n+3$ labels the sections of the $\left(\mbf{1},\mbf{3}\right)$- and $\left(\mbf{3},\mbf{1}\right)$-bundles. These sections thus satisfy
\begin{equation}
 \omega_{M,ab} n^b = 0 \,, \label{eq:omegaConstraint}
\end{equation}
and we further normalise them according to
\begin{equation}
 n^a n_a = 1 \,, \qquad \omega_{M,ab} \omega_{N,cd} \epsilon^{abcde} = 4 \eta_{MN} n^e \,. \label{eq:SectionNormalise}
\end{equation}
Here $\eta_{MN}$ has signature $\left(3,n\right)$ reflecting the number of sections of the $\left(\mbf{1},\mbf{3}\right)$- and $\left(\mbf{3},\mbf{1}\right)$-bundles used.

It is useful to also introduce
\begin{equation}
 \omega_M{}^{ab} = \epsilon^{abcde} \omega_{M,cd} n_e \,,
\end{equation}
which satisfy
\begin{equation}
 \omega_M{}^{ab} \omega_{N,ab} = 4 \eta_{MN} \,, \qquad \omega_M{}^{ab} n_b = 0 \,,
\end{equation}
as a result of \eqref{eq:SectionNormalise} and \eqref{eq:omegaConstraint}. Further useful identities are
\begin{equation}
 \begin{split}
  \omega_{(M}{}^{cd} \omega_{N)ca} &= \eta_{MN} \left( \delta_a{}^b - n_a n^b \right) \,, \\
  \omega_{M,ab} \epsilon^{abcde} &= 3 \omega_M{}^{[cd} n^{e]} \,, \\
  \omega_M{}^{ab} \epsilon_{abcde} &= 12 \omega_{M[cd} n_{e]} \,, \\
  \omega_{M}{}^{ab} \omega_N{}^{cd} \epsilon_{abcde} &= 16 \eta_{MN} n^e \,.
 \end{split} \label{eq:OmegaIdentities}
\end{equation}
We will also often make use of the generalised vector
\begin{equation}
 \tomega_M{}^{ab} = \rho \omega_M{}^{ab} \,,
\end{equation}
which has weight $\frac15$ and thus can be used as a generator of generalised diffeomorphisms.

\subsubsection{Truncation Ansatz} \label{s:TruncAnsatz}
One can now perform a truncation Ansatz by expanding all the fields of the $\SL{5}$ EFT in terms of $\rho$, $n_a$, $n^a$ and $\omega_{M,ab}$. We will label the truncation Ansatz by the brackets $\langle \rangle$. For the scalar fields $\kappa$, $A_a$, $A^a$ and $B_{u,ab}$ it is given by
\begin{equation}
 \begin{split}
  \langle\kappa\rangle(x,Y) &= |\bae|^{1/7}(x)\,e^{-2d(x)/5}\, \rho(Y) \,, \\
  \langle A^a\rangle(x,Y) &= \frac{1}{\sqrt{2}} e^{-4d(x)/5} n^a(Y) \,, \\
  \langle A_a\rangle(x,Y) &= \frac{1}{\sqrt{2}} e^{4d(x)/5} n_a(Y) \,, \\
  \langle B_{u,ab} \rangle(x,Y) &= e^{-2d(x)/5}\, b_{u,M}(x) \omega^M{}_{ab}(Y) \,,
\label{eq:ScalarAnsatz}
 \end{split}
\end{equation}
and hence
\begin{equation}
 \langle V_u{}^{ab} \rangle = \frac{1}{\sqrt{2}} e^{2d(x)/5}\, b_{u,M}(x)\, \omega^{M,ab}(Y) \,.
\end{equation}

In order to satisfy the compatibility conditions \eqref{eq:CompatibilityConditions}, the $b_{u,M}$ are subject to the constraint
\begin{equation}
 b_{u,M} b_{v,N} \eta^{MN} = \delta_{uv} \,. \label{eq:bConstraint}
\end{equation}
Furthermore, the $u$ index labels the triplet of $\SU{2}_R$ and we wish to identify any objects related by $R$-symmetry. This leaves $3n$ degrees of freedom in $b_{u,M}$ which is also the dimension of the coset space $\frac{ON{3,n}}{\ON{3}\times\ON{n}}$. Indeed, we can use $b_{u,M}$ to define a symmetric group element of $\ON{3,n}$ by
\begin{equation}
 b_{u,M} b^u{}_N = \frac12 \left( \eta_{MN} - \gH_{MN} \right) \,.
\end{equation}
Because of \eqref{eq:bConstraint}, $\gH_{MN}$ satisfies
\begin{equation}
 \gH_{MP} \gH_{NQ} \eta^{PQ} = \eta_{MN} \,,
\end{equation}
showing that it is an element of $\ON{3,n}$. $\gH_{MN}$ can be identified as the generalised metric of seven-dimensional gauged SUGRA and we will see here that it also becomes the generalised metric of the heterotic DFT. Additionally, the scalars $|\bae|$ will become the determinant of the external seven-dimensional vielbein and $d$ the generalised dilaton of the heterotic DFT.

The truncation Ans\"atze for the remaining fields are
\begin{equation}
 \begin{split}
  \langle \TA_\mu{}^{ab} \rangle (x,Y) &= A_\mu{}^M(x)\, \omega_M{}^{ab}(Y) \, \rho(Y) \,, \\
  \langle \TB_{\mu\nu,a} \rangle (x,Y) &= -4 B_{\mu\nu}(x) \, n_a(Y) \, \rho^2(Y) \,, \\
  \langle \TC_{\mu\nu\gamma}{}^a \rangle (x,Y) &= C_{\mu\nu\gamma}(x) \, n^a(Y) \, \rho^3(Y) \,, \\
  \langle \TD_{\mu\nu\gamma\sigma,ab} \rangle(x,Y) &= D_{\mu\nu\gamma\sigma\,M}(x)\, \omega^{M}{}_{ab}(Y)\, \rho^4(Y) \,, \\
  \langle e_\mu{}^{\bmu} \rangle (x,Y) &= \bae_\mu{}^{\bmu}(x)\, e^{-2d(x)/5}\, \rho(Y) \,.
 \end{split} \label{eq:GaugeMetricAnsatz}
\end{equation}

\subsubsection{Consistency conditions and embedding tensor} \label{s:TruncConsistencyConditions}
In order to have a consistent truncation one needs to impose a set of differential constraints on the section $\rho$, $n_a$, $n^a$ and $\omega_{M,ab}$ which define the truncation. The so-called ``doublet'' conditions
\begin{equation}
 \begin{split}
  n^a \gL_{\tomega_M} \tomega_{N,ab} &= 0 \,, \\
  \gL_{\tomega_M} n^a &= n^a n_b \gL_{\tomega_M} n^b \,, \\
  \partial_{ab} \left( n^b \rho^3 \right) &= \rho^3 n_a n^b \partial_{bc} n^c \,,
 \end{split} \label{eq:DoubletConditions}
\end{equation}
ensure that the $\left(\mbf{2},\mbf{2}\right)$ and $\left(\mbf{2},\mbf{4}\right)$ representation of the intrinsic torsion vanish. This is required in order to avoid couplings to $\SU{2}_S$ doublets in the $\SL{5}$ fields which we want to remove in the truncation Ansatz in order to have a ${\cal N}=2$ theory.

In addition, we require the sections $\omega_{M,ab}$ to form a closed set under the generalised Lie derivative, i.e.
\begin{equation}
 \gL_{\tomega_M} \omega^N{}_{ab} = \frac14 \left( \gL_{\tomega_M} \omega^N{}_{cd} \right) \omega_P{}^{cd} \omega^P{}_{ab} \,. \label{eq:ClosureCondition}
\end{equation}

Given the \eqref{eq:DoubletConditions} and \eqref{eq:ClosureCondition} one can identify the object
\begin{equation}
 g_{MNP} \equiv \frac14 \gL_{\tomega_M} \omega_{N,ab} \omega_{P}{}^{ab} \,, \label{eq:TotalFluxes}
\end{equation}
with the embedding tensor of the half-maximal gauged SUGRA. In particular, it has only three irreducible representations, given by two $\ON{n+3}$ vectors
\begin{equation}
 f_M = n^a \gL_{\tomega_M} n_a \,, \qquad \xi_M = \rho^{-1} \gL_{\tomega_M} \rho \,, \label{eq:VectorFluxes}
\end{equation}
and a totally antisymmetric 3-index tensor
\begin{equation}
 f_{MNP} = g_{[MNP]} \,, \label{eq:EmbeddingTensor}
\end{equation}
which are the only representations allowed by the linear constraint of seven-dimensional half-maximal gauged SUGRA \cite{Dibitetto:2012rk}. One can also identify the $p=3$ deformation \cite{Bergshoeff:2007vb} with
\begin{equation}
 \Theta = \rho n^a \partial_{ab} n^b \,. \label{eq:SingletFlux}
\end{equation}

It is easy to see that just as in the maximal case, the closure of the algebra of generalised diffeomorphisms implies the quadratic constraint of the gauged SUGRA. Thus, imposing the section condition on $\rho$, $n_a$, $n^a$ and $\omega_{M,ab}$ is sufficient to satisfy the quadratic constraints of gauged SUGRA. Additionally, if one wants to obtain an action principle, one must ensure that $\xi_M$, the so-called trombone tensor, vanishes. Using the truncation Ansatz \eqref{eq:ScalarAnsatz}, \eqref{eq:GaugeMetricAnsatz} one finds that the dependence on the internal coordinates, $Y^{ab}$, appears only through the embedding tensor and as an overall factor of $\rho^5$. Thus, we obtain a consistent truncation when the sections obey \eqref{eq:DoubletConditions}, \eqref{eq:ClosureCondition} and the embedding tensor given \eqref{eq:EmbeddingTensor}, \eqref{eq:VectorFluxes} and \eqref{eq:SingletFlux} is constant and obeys the quadratic constraint.

\section{Heterotic DFT as a half-maximal reduction of EFT}\label{s:HetDeformation}

We will now show that the above set-up can be used to reduce the $\SL{5}$ EFT to the $\ON{3,n}$ heterotic DFT. This requires half of the supersymmetry to be broken which can be achieved by reducing the theory on a generalised $\SU{2}$-structure manifold living in the extended space. In particular, we will use the Ans\"atze \eqref{eq:ScalarAnsatz} and \eqref{eq:GaugeMetricAnsatz} but we will allow the coefficients, which in the truncation Ansatz only depend on the seven $x^\mu$ coordinates, to also depend on the extended space $Y^{ab}$. These coefficients will then become the heterotic DFT fields.

However, for consistency we will have to impose certain restrictions of their the dependence on the $Y^{ab}$. We will see that these conditions allow us to define the $n+3$ internal derivatives of the heterotic DFT by ``twisting'' the 10 extended coordinate derivatives $\partial_{ab}$ and that the generalised Lie derivative of the $\SL{5}$ EFT will reduce to the heterotic generalised Lie derivative. The gauge group of the heterotic DFT will be determined by the embedding tensor defined by the $\SU{2}$-structure reduction.

What we are doing here is reminiscent of the procedure to relate EFT to massive IIA theory \cite{Ciceri:2016dmd}. There a reduction was performed on a twisted torus in the extended space and the resulting reduced EFT fields had a limited dependence on the $Y^{ab}$ which upon solving the section condition led to massive IIA theory (or its IIB dual).

\subsection{Reduction Ansatz}\label{s:HetAnsatz}
We will now use the same Ansatz for the scalars, gauge fields and spinors as in the consistent ${\cal N}=2$ truncation Ansatz of \cite{Malek:2016bpu}, i.e. equations \eqref{eq:ScalarAnsatz}, \eqref{eq:GaugeMetricAnsatz}, but still allow the coefficients to depend on some of the extended space coordinates $Y^{ab}$. Thus we now write the scalar fields as
\begin{equation}
 \begin{split}
  \langle\kappa\rangle(x,Y) &= |\bae(x,Y)|^{-1/7}\, e^{-2d(x,Y)/5}\, \rho(Y) \,, \\
  \langle A^a\rangle(x,Y) &= \frac{1}{\sqrt{2}} e^{-4d(x,Y)/5}\, \rho(Y)\, n^a(Y) \,, \\
  \langle A_a\rangle(x,Y) &= \frac{1}{\sqrt{2}} e^{4d(x,Y)/5}\, \rho(Y)\, n_a(Y) \,, \\
  \langle B_{u,ab} \rangle(x,Y) &= e^{-2d(x,Y)/5}\, b_{u,M}(x,Y)\, \omega^M{}_{ab}(Y) \, \rho(Y) \,.
 \end{split}
\end{equation}
and for the gauge fields and external vielbein
\begin{equation}
 \begin{split}
  \langle \TA_\mu{}^{ab} \rangle (x,Y) &= A_\mu{}^M(x,Y)\, \omega_M{}^{ab}(Y) \, \rho(Y) \,, \\
  \langle \TB_{\mu\nu,a} \rangle (x,Y) &= -4 B_{\mu\nu}(x,Y) \, n_a(Y) \, \rho^2(Y) \,, \\
  \langle \TC_{\mu\nu\gamma}{}^a \rangle (x,Y) &= C_{\mu\nu\gamma}(x,Y) \, n^a(Y) \, \rho^3(Y) \,, \\
  \langle \TD_{\mu\nu\gamma\sigma\,ab} \rangle (x,Y) &= D_{\mu\nu\gamma\sigma\,M}(x,Y) \omega^M{}_{ab}(Y) \, \rho^4(Y) \,, \\
  \langle e_\mu{}^\bmu \rangle(x,Y) &= \bae_{\bmu}{}^\mu(x,Y)\, e^{-2d(x,Y)/5}\, \rho(Y) \,,
 \end{split}
\end{equation}
where will show that $\bae_{\bmu}{}^\mu$ is the string-frame vielbein of the heterotic DFT, and the factor of $-4$ allows us to make contact with the conventions in \cite{Hohm:2013nja}.

Exactly as for the truncation Ansatz we require the sections $\rho$, $n^a$, $n_a$ and $\omega_{M,ab}$ to satisfy the same consistency condition as those required for consistent truncations, see section \ref{s:ConsTruncation}. Therefore we impose equations \eqref{eq:ClosureCondition} and \eqref{eq:DoubletConditions} and require that the embedding tensor, given in equations \eqref{eq:EmbeddingTensor} and \eqref{eq:VectorFluxes}, as well as the singlet deformations \eqref{eq:SingletFlux} are constant and satisfy the quadratic constraint. To allow for a close comparison to the heterotic DFT formulation in \cite{Siegel:1993th,Siegel:1993xq,Hohm:2011ex} we will take $f_M = \xi_M = \Theta = 0$. It is however easy to include these, although when $\xi_M \neq 0$ we will not obtain a consistent action principle.

However, let us emphasise that we are not performing a truncation because the fields $\bae$, $d$, $b_u{}^M$, $A_\mu{}^M$, $B_{\mu\nu}$, $C_{\mu\nu\rho}$ are still allowed to depend on $Y^{ab}$. Instead, we are performing a reduction of the theory, which as we will see produces the heterotic $\SO{3,n}$ DFT with a $7+3$ split. This is similar to the procedure used in \cite{Ciceri:2016dmd} to obtain massive IIA from EFT.

Recall that the compatibility condition \eqref{eq:CompatibilityConditions} implies that
\begin{equation}
 b_{u,M} b_v{}^M = \delta_{uv} \,,
\end{equation}
and this allows us to define the generalised metric as
\begin{equation}
 \gH_{MN} = 2 b_{u,M} b^u{}_N - \eta_{MN} \,.
\end{equation}
We see that the $b_{u,M}$ appear exactly like the frame fields in DFT \cite{Hohm:2010pp,Hohm:2011ex}. We will see in section \ref{s:HetPotential} that indeed the frame formulation appears naturally from the $\SU{2}$-reduction of EFT. We will also make use of the left- and right-moving projectors
\begin{equation}
 \begin{split}
  P^-_{MN} &= b_{u,M} b^u{}_N = \frac12 \left( \eta_{MN} - \gH_{MN} \right) \,, \\
  P^+_{MN} &= \eta_{MN} - b_{u,M} b^u{}_N = \frac12 \left( \eta_{MN} + \gH_{MN} \right) \,.
 \end{split}
\end{equation}

\subsection{Doublet and closure conditions}
In addition to the differential conditions imposed on the sections $\rho$, $n_a$, $n^a$ and $\omega_{M,ab}$ we must also impose certain differential constraints on the fields of the reduced theory because these now carry dependence on the extended space.

As discussed in \ref{s:ConsTruncation} and in more length in \cite{Malek:2016bpu}, we must project out the doublets of $\SU{2}_S$ in order to have a ${\cal N}=2$ theory. Here this now means that we require that the fields of the reduced EFT do not depend on the doublet coordinates, i.e.
\begin{equation}
 n^a \partial_{ab} d = n^a \partial_{ab} a = n^a \partial_{ab} b_{u,M} = \ldots = 0 \,.
\end{equation}
This removes the dependence on four coordinates so that the reduced theory is left with a six-dimensional extended coordinate space.

Furthermore, we require that the remaining dependence on these $\left(\mbf{1},\mbf{3}\right) \oplus \left(\mbf{3},\mbf{1}\right)$ coordinates can be expanded in terms of the sections $\omega_M{}^{ab}$, i.e.
\begin{equation}
 \partial_{ab} = \frac14 \omega_{ab}{}^{M} \omega_M{}^{cd} \partial_{cd} \,,
\end{equation}
when acting on any of the fields in the reduced theory, e.g. for a vector $V^M$
\begin{equation}
 \partial_{ab} V^M = \frac14 \omega_{ab}{}^M \omega_M{}^{cd} \partial_{cd} V^M \,.
\end{equation}

\subsection{Reduced extended space and $\ON{3,n}$ section condition} \label{s:HetSpace}
Because the derivatives acting on the reduced fields can be expanded in the $\omega_{M,ab}$'s we can introduce the ``twisted'' derivatives
\begin{equation}
 D_M V^N = \frac12 \rho\, \omega_M{}^{cd} \partial_{cd} V^N \,,
\end{equation}
and as we will see these will become the $n+3$ derivatives of the extended heterotic space as in \cite{Hohm:2011ex}. In order for this identification to work, we require their commutator to vanish, i.e.
\begin{equation}
 \left[ D_M, D_N \right] V^P = 0 \,,
\end{equation}
for a generic object $V^P$ of the reduced theory. We can write
\begin{equation}
 \left[ D_M, D_N \right] = \frac1{2\rho} \gL_{\tomega_M} \tomega_N{}^P D_P - \frac32 \tomega_N{}^{[ab} \partial_{ab} \tomega_M{}^{cd]} \partial_{cd} \,,
\end{equation}
and using the result of section \ref{s:TruncConsistencyConditions} we find
\begin{equation}
 \left[ D_M, D_N \right] = f_{MN}{}^P D_P - \frac32 \tomega_N{}^{[ab} \partial_{ab} \tomega_M{}^{cd]} \partial_{cd} \,.
\end{equation}
The first term is thus proportional to the embedding tensor of the extended space while the second term is proportional to the section condition acting on the background and the objects of the reduced theory. We now impose both conditions separately, i.e.
\begin{equation}
 \tomega_N{}^{ab} \partial_{[ab} \tomega_{|M|}{}^{cd} \partial_{cd]} V^N = 0 \,,
\end{equation}
as well as
\begin{equation}
 f_{MN}{}^P D_P V^Q = 0 \,,
\end{equation}
where as before $V^M$ denotes a generic object of the reduced theory. Given these conditions the derivatives $D_M$ commute and thus we can treat them as if they were $(n+3)$ partial derivatives. The condition
\begin{equation}
 f_{MN}{}^P \partial_P V^N = 0 \,, \label{eq:GaugeSC}
\end{equation}
is precisely the condition imposed in heterotic DFT, where $f_{MN}{}^P$ is the embedding tensor encoding the gauge group of the heterotic supergravity \cite{Hohm:2011ex}. Thus we see the first evidence that the embedding tensor of the $\SU{2}$-structure reduction defines the gauge group of the heterotic DFT.

We should mention that in the case of massive IIA \cite{Ciceri:2016dmd} a similar analysis is used to show that the reduced theory has a more restricted ``section condition''. This implies that the reduced theory can only contain 10-dimensional solutions, not 11-dimensional ones, as we expect for a theory with a Roman's mass parameter.

Returning to the $\SU{2}$-reduction, it may at first seem strange that we can treat the $\left(n+3\right)$ derivatives as if they were coordinate derivatives $\partial_M$ even though we started off with only a 10-dimensional extended space. How can all the $\left(n+3\right)$ derivatives be independent? The answer is of course that they are not but when acting on fields in the reduced theory they can be treated as such because the fields do not have arbitrary coordinate dependence. Their coordinate dependence is restricted by the EFT section condition which we have not yet imposed. This now takes the form
\begin{equation}
 \partial_{[ab} V^P \partial_{cd]} W^Q = \frac{1}{3!} \rho^{-2} \epsilon_{abcde} \eta^{MN} n^e \partial_M V^P \partial_N W^Q = 0 \,,
\end{equation}
and similarly for double derivatives, where $V^P$ and $W^Q$ represent arbitrary $\ON{3,n}$ fields. Thus we obtain the $\ON{3,n}$ section condition
\begin{equation}
 \eta^{MN} \partial_M f \partial_N g = \eta^{MN} \partial_M \partial_N f = 0 \,, \label{eq:SO3nSC}
\end{equation}
for any fields $f$ and $g$ of the reduced theory. This implies that the reduced fields can only depend on three coordinates and thus although we formally use the objects $\partial_M$, only three of these are ever non-zero.

\subsection{Heterotic generalised Lie derivative}\label{s:HetLieDerivative}
Let us now show that with the Ansatz described in section \ref{s:HetAnsatz} above, the EFT generalised Lie derivative reduces to the heterotic DFT generalised Lie derivative.

We wish to calculate the generalised Lie derivative of two generalised vectors
\begin{equation}
 \langle \Lambda^{ab} \rangle = \Lambda^M(x,Y) \rho(Y) \, \omega_M{}^{ab}(Y) \,,  \qquad \langle V^{ab} \rangle = V^M(x,Y) \rho(Y)\, \omega_M{}^{ab}(Y) \,,
\end{equation}
where as discussed we now let $\Lambda^{M}(x,Y)$ and $V^M(x,Y)$ only depend on the external seven coordinates $x^\mu$ and the $\left(\mbf{3},\mbf{1}\right) \oplus \left(\mbf{1},\mbf{3}\right)$ extended coordinates. We then find
\begin{equation}
 \begin{split}
  \langle \gL_\Lambda V^{ab} \rangle &= \frac12 \Lambda^M \tomega_M{}^{cd} \tomega_N{}^{ab} \partial_{cd} V^N + \frac12 V^N \tomega_N{}^{ab} \tomega_M{}^{cd} \partial_{cd} \Lambda^M - 2 V^N \tomega_N{}^{c[b} \tomega_M{}^{a]d} \partial_{cd} \Lambda^M \\
  & \quad + \Lambda^M V^N \gL_{\tomega_M} \tomega_N{}^{ab} \\
  &= \bar{\gL}_{\Lambda} V^{ab} + f_{MN}{}^P V^M \Lambda^N \tomega_P{}^{ab} \,,
 \end{split}
\end{equation}
where we have used \eqref{eq:EmbeddingTensor} to write the final term as the embedding tensor defined by the $\SU{2}$-structure, and we have defined
\begin{equation}
 \begin{split}
  \bar{\gL}_{\Lambda} V^{ab} &= \tomega_M{}^{ab} \left( \frac12 \Lambda^N \tomega_N{}^{cd} \partial_{cd} V^M + \frac12 V^M \tomega_N{}^{cd} \partial_{cd} \Lambda^N \right) - 2 V^N \tomega_N{}^{c[b} \tomega_M{}^{a]d} \partial_{cd} \Lambda^M \,.
 \end{split}
\end{equation}
The first two terms are just
\begin{equation}
 \frac12 \Lambda^N \tomega_N{}^{cd} \partial_{cd} V^M + \frac12 V^M \tomega_N{}^{cd} \partial_{cd} \Lambda^N = \Lambda^N \partial_N V^M + V^M \partial_N \Lambda^N \,,
\end{equation}
where as we discussed above we defined
\begin{equation}
 \partial_M = \frac12 \rho \, \omega_M{}^{ab} \partial_{ab} \,.
\end{equation}
For the last term we use \eqref{eq:OmegaIdentities} to write
\begin{equation}
 \omega^M{}_{ab} \omega_N{}^{ca} \omega_P{}^{bd} = - 2 \omega_N{}^{cd} \delta_P^M + \eta_{NP} \omega^{M,bc} \,.
\end{equation}
Hence we find
\begin{equation}
 \begin{split}
  \bar{\gL}_{\Lambda} V^{ab} &= \tomega_M{}^{ab} \left( \Lambda^N \partial_N V^M + V^M \partial_N \Lambda^N - 2 V^{(M} \partial_N \Lambda^{N)} + \eta^{MP} \eta_{NQ} V^N \partial_P \Lambda^Q \right) \\
  &= \tomega_M{}^{ab} \bar{\gL}_{\Lambda} V^M \,,
 \end{split}
\end{equation}
where
\begin{equation}
 \bar{\gL}_{\Lambda} V^M = \Lambda^N \partial_N V^M - V^N \partial_N \Lambda^M + \eta^{MP} \eta_{NQ} V^N \partial_P \Lambda^Q \,,
\end{equation}
is precisely the $\SO{3,n}$ generalised Lie derivative with no gauging. Putting everything together we obtain
\begin{equation}
 \langle \gL_{\Lambda} V^{ab} \rangle = \tomega_{M}{}^{ab} L_\Lambda V^M \,,
\end{equation}
where we defined
\begin{equation}
 L_\Lambda V^M = \Lambda^N \partial_N V^M - V^N \partial_N V^M + \eta^{MP} \eta_{NQ} V^N \partial_P \Lambda^Q + f_{NP}{}^M V^N \Lambda^P \,. \label{eq:HetLie}
\end{equation}
This is the generalised Lie derivative of the heterotic DFT, with gauge group defined by the embedding tensor $f_{NP}{}^M$, see equation (3.6) of \cite{Hohm:2011ex}.

\section{Heterotic Action} \label{s:HetAction}
We will now show how the EFT action reduces to that of the heterotic DFT \cite{Hohm:2011ex} with a Kaluza-Klein split with seven external dimensions \cite{Hohm:2013nja}.

\subsection{Heterotic scalar potential}\label{s:HetPotential}
Using the Ansatz \eqref{eq:ScalarAnsatz} it is easy to show that the EFT scalar potential reduces to the scalar potential of the heterotic DFT. Let us begin by evaluating the intrinsic torsion with the Ansatz \eqref{eq:ScalarAnsatz} and using \eqref{eq:EmbeddingTensor}, \eqref{eq:VectorFluxes} and \eqref{eq:SingletFlux}. As we discussed earlier we take $f_M = \xi_M = \Theta = 0$ to allow for a close comparison with \cite{Hohm:2011ex}. We find that the doublets of $\SU{2}_S$ vanish exactly as required. The remaining irreducible representations become
\begin{equation}
 \begin{split}
  \langle S \rangle &= \frac1{2\rho} e^{-8d/5} \Theta \,, \\
  \langle T \rangle &= \frac1{6 \rho} e^{2d/5} \epsilon^{uvw} \Omega_{uvw} \,, \\
  \langle T_u \rangle &= \frac{1}{\rho\sqrt{2}} e^{2d/5} \left[ \Omega_u + \frac47 b_u{}^M \partial_M \ln|e| \right] \,, \\
  \langle S_u \rangle &= \frac{\sqrt{2}}{\rho} e^{2d/5} \left[ \Omega_u + \frac67 b_u{}^M \partial_M \ln|e| \right] \,, \\
  \langle T_{ab} \rangle &= \frac1{28\rho\sqrt{2}}\, P_+^M{}_N \omega^N{}_{ab} \partial_{M} \ln|e| \,, \\
  \langle T^u{}_{ab} \rangle &= \frac{1}{12\rho\sqrt{2}}\, \epsilon^{uvw} e_{\bar{u}}{}^{M} \omega_{M,ab} \Omega_{vw}{}^{\bar{u}} \,.
 \end{split} \label{eq:HetTTensor}
\end{equation}
Here we have defined the generalised coefficients of anholonomy $\Omega_{uvw}$ as in \cite{Hohm:2011ex} (see also \cite{Siegel:1993xq},\cite{Hohm:2010xe}, \cite{Hohm:2010pp}). That is,
\begin{equation}
 \Omega_{uvw} = \left( L_{b_{[u}} b_v{}^N \right) b_{w]N} \,,
\end{equation}
in terms of the generalised Lie derivative of the heterotic DFT including the gaugings, i.e. \eqref{eq:HetLie}, and
\begin{equation}
 \Omega_u = \partial_M b_u{}^M - 2 b_u{}^M \partial_M d \,.
\end{equation}

Furthermore, $e_{\bar{u}}{}^M$, with $\bar{u} = 1, \ldots, n$ represent the $n$ right-moving vielbeine of the generalised metric satisfying
\begin{equation}
 P_+^M{}_N e_{\bar{u}}{}^N = e_{\bar{u}}{}^M \,, \qquad P_+^{MN} = e_{\bar{u}}{}^M e^{\bar{v}N} \eta^{\bar{u}\bar{v}} \,,
\end{equation}
with $\eta^{\bar{u}\bar{v}}$ an $\ON{n}$ metric which is not necessarily constant, and $\Omega_{uv}{}^{\bar{u}}$ is similarly defined in terms of these generalised vielbeins, see \cite{Hohm:2011ex}.

Plugging \eqref{eq:HetTTensor} into \eqref{eq:RicciScalar} and using the Ansatz for the metric \eqref{eq:GaugeMetricAnsatz} in \eqref{eq:Nablag} we find the heterotic DFT scalar potential in the frame formulation \cite{Hohm:2011ex}.
\begin{equation}
 \begin{split}
  \langle |e| V \rangle &= \rho^5 |\bae| e^{-2d} \left[ \frac{1}{12} \Omega_{uvw} \Omega^{uvw} + \frac14 \eta_{\bar{u}\bar{v}} \Omega_{uv}{}^{\bar{u}} \Omega^{uv\bar{v}} + \frac12 \Omega_u \Omega^u + \Omega^u b_u{}^M \partial_M \ln |e| \right. \\
  & \left. - \frac14 \gH^{MN} \partial_M g^{\mu\nu} \partial_N g_{\mu\nu} \right] \,.
 \end{split}
\end{equation}

\subsection{Heterotic kinetic and topological terms}\label{s:HetKinTerms}
We have already shown in section \ref{s:HetLieDerivative} that the EFT generalised Lie derivative reduces to that of the heterotic DFT. This means that the external covariant derivative of the EFT \cite{Hohm:2013uia}, defined as
\begin{equation}
 D_\mu = \partial_\mu - \gL_{A_\mu}\,,
\end{equation}
will also reduce to the heterotic external covariant derivative, see e.g. \cite{Hohm:2013nja} for the $\ON{d,d}$ version thereof.

For example, we find
\begin{equation}
 \begin{split}
  \langle D_\mu  B_{u,ab} \rangle &= \omega_{M,ab} \left[ \partial_\mu \left( b_u{}^M e^{-2d/5} \right) - L_{A_\mu} \left( b_u{}^M e^{-2d/5} \right) \right] \,, \\
  \langle D_\mu A^a \rangle &= \frac{1}{\sqrt{2}} n^a \left( \partial_\mu e^{-4d/5} - L_{A_\mu} e^{-4d/5} \right) \,, \\
  \langle D_\mu \kappa \rangle &= \rho \left[ \partial_\mu \left( |\bae|^{1/7} e^{-2d/5} \right) - L_{A\mu} \left( |\bae|^{1/7} e^{-2d/5} \right) \right] \,,
 \end{split}
\end{equation}
where one can read off
\begin{equation}
 \begin{split}
  L_{A_\mu} \left( e^{-2d/5} b_u{}^M \right) &= A_\mu{}^N \partial_N \left( e^{-2d/5} b_u{}^M \right) - e^{-2d/5} b_u{}^N \partial_N A_\mu{}^M + e^{-2d/5} b_u{}^N \partial^M A_\mu{}^N \\
  & \quad+ \frac15 e^{-2d/5} b_u{}^M \partial_N A_\mu{}^N + f_{NP}{}^M A_\mu{}^N b_u{}^P \,, \\
  L_{A_\mu} e^{-4d/5} &= A_\mu{}^N \partial_N e^{-4d/5} + \frac25 e^{-4d/5} \partial_N A_\mu{}^N \,, \\
  L_{A_\mu} \left( |\bae|^{1/7} e^{-2d/5} \right) &= A_\mu{}^N \partial_N \left( |\bae|^{1/7} e^{-2d/5} \right) + \frac15 |\bae|^{1/7} e^{-2d/5} \partial_N A_\mu{}^N \,.
 \end{split}
\end{equation}

These equations imply that
\begin{equation}
 \begin{split}
  L_{A_\mu} b_u{}^M &= A_\mu{}^N \partial_N b_u{}^M - b_u{}^N \partial_N A_\mu{}^M + b_u{}^N \partial^M A_\mu{}^N + f_{NP}{}^M A_\mu{}^N b_u{}^P \,, \\
  L_{A_\mu} e^{-2d} &= A_\mu{}^N \partial_N e^{-2d} + e^{-2d} \partial_N A_\mu{}^N \,, \\
  L_{A_\mu} |\bae| &= A_\mu{}^N \partial_N |\bae| \,.
 \end{split}
\end{equation}
Thus, we can see that $b_u{}^M$ transforms as a $\ON{3,n}$ vector field, $e^{-2d}$ as a scalar density of weight $1$ and $|\bae|$ as a scalar, exactly as in the heterotic DFT.

The same computation for the full external vielbein and the gauge fields shows that $\bae_\mu{}^{\bmu}$ transforms as a scalar, $A_\mu{}^M$ as a $\ON{3,n}$ vector and $B_{\mu\nu}$ and $C_{\mu\nu\rho}$ as scalars with respect to the heterotic generalised Lie derivative. Thus we see that the EFT covariant derivative $D_\mu$ reduces to that of the heterotic DFT which we label by
\begin{equation}
 \D_\mu = \partial_\mu - L_{A_\mu} \,.
\end{equation}

We can now compute
\begin{equation}
 \begin{split}
  \langle D_\mu A^a \rangle &= -\frac{4}{5\sqrt{2}} e^{-4d/5} n^a \D_\mu d \,, \\
  \langle \left( D_\mu B_{u,ab} D_\nu B^u{}_{cd} \right) \epsilon^{abcde} A_e \rangle &= 2 \sqrt{2} \D_\mu b_u{}^M \D_\nu b^u{}_M + \frac{24\sqrt{2}}{25} \D_\mu d \D_\nu d \,. \label{eq:HetDmuAB}
 \end{split}
\end{equation}
Using
\begin{equation}
 \partial_\mu \gH^{MN} \partial_\nu \gH_{MN} = 8 \partial_\mu b_u{}^N \partial_\nu b^u{}_N \,,
\end{equation}
we can rewrite the second equation as
\begin{equation}
 \langle \left( D_\mu B_{u,ab} D_\nu B^u{}_{cd} \right) \epsilon^{abcde} A_e \rangle = \frac{\sqrt{2}}{4} \D_\mu \gH^{MN} \D_\nu \gH_{MN} + \frac{24\sqrt{2}}{25} \D_\mu d \D_\nu d \,. \label{eq:HetDmuB}
\end{equation}
Thus, the scalar kinetic terms reduce to
\begin{equation}
 \begin{split}
  \langle |e| L_{\textrm{SK}} \rangle &= \langle |e| g^{\mu\nu} \left( \frac{1}{2\sqrt{2}} D_\mu B_{u,ab} D_\nu B^u{}_{cd} \epsilon^{abcde} A_e - 14 D_\mu A^a D_\nu A_a \right) \rangle \\
  &= \rho^{5} |\bae| e^{-2d} \left( \frac18 \bag^{\mu\nu} \D_\mu \gH^{MN} \D_\nu \gH_{MN} + 4 \bag^{\mu\nu} \D_\mu d\, \D_\nu d \right) \,,
 \end{split}
\end{equation}
which are the the scalar kinetic terms of the heterotic DFT in a Kaluza-Klein split, see e.g. \cite{Hohm:2013nja}.

Let us now consider the reduction of the field strengths using \eqref{eq:GaugeMetricAnsatz}. We find
\begin{equation}
 \begin{split}
  \langle \Fa_{\mu\nu}{}^{ab} \rangle &= \rho\, \omega_M{}^{ab} F_{\mu\nu}{}^M \,, \\
  \langle \Fb_{\mu\nu\gamma\,a} \rangle &= - 4 \rho^2 n_a H_{\mu\nu\gamma} \,, \\
  \langle \Fc_{\mu\nu\gamma\sigma}{}^a \rangle &= \rho^3 n^a J_{\mu\nu\gamma\sigma} \,,
 \end{split}
\end{equation}
where $F_{\mu\nu}{}^M$, $H_{\mu\nu\gamma}$ and $J_{\mu\nu\gamma\sigma}$ are the reduced field strength of the Kaluza-Klein split heterotic DFT \cite{Hohm:2013nja}, i.e.
\begin{equation}
 \begin{split}
  F_{\mu\nu}{}^M & = 2 \partial_{[\mu} A_{\nu]}{}^M - \left[ A_\mu, A_\nu \right]_C^M - \partial^M B_{\mu\nu} \,, \\
  H_{\mu\nu\rho} &= 3 \D_{[\mu} B_{\nu\rho]} + 3 \partial_{[\mu} A_\nu{}^M A_{\rho]M} - A_{[\mu}{}^M \left[ A_\nu, A_{\rho]} \right]_{C,M} \,, \\
  J_{\mu\nu\rho\sigma} &= 4 \D_{[\mu} C_{\nu\rho\sigma]} + \partial^M D_{\mu\nu\rho\sigma,M} \,.
 \end{split}
\end{equation}
Here
\begin{equation}
 \left[ A_\mu, A_\nu \right]_C^M = \frac12 \left( L_{A_\mu} L_{A_\nu} - L_{A_\nu} A_\mu \right) \,,
\end{equation}
is the antisymmetrised heterotic generalised Lie derivative. Because the three form decouples from the two-form, it is not necessary to include it in the Kaluza-Klein split DFT tensor hierarchy.

With the above reduction it is easy to see that
\begin{equation}
 \begin{split}
  \langle |e| L_{\textrm{kin,vectors}} \rangle &= \rho^{5} |\bae| e^{-2d} \bag^{\mu\gamma} \bag^{\nu\sigma} F_{\mu\nu}{}^M F_{\gamma\sigma}{}^N \left( 2 b_{u,M} b^u{}_N - \eta_{MN} \right) \\
  &= - \rho^{5} |\bae| e^{-2d} \bag^{\mu\gamma} \bag^{\nu\sigma} F_{\mu\nu}{}^M F_{\gamma\sigma}{}^N \gH_{MN} \,,
 \end{split}
\end{equation}
which is the correct kinetic term for the vector fields. Similarly, we reduce the kinetic term for the two-form potentials
\begin{equation}
 L_{kin,\textrm{2-form}} = -\frac{1}{48} \Fb_{\mu\nu\rho,a} \Fb^{\mu\nu\rho}{}_b A^a A^b \,,
\end{equation}
to find
\begin{equation}
 \langle |e| L_{\textrm{kin,2-form}} \rangle = - \frac{1}{6} \rho^{5} |\bae| e^{-2d} \bag^{\mu\sigma} \bag^{\nu\rho} \bag^{\gamma\lambda} H_{\mu\nu\gamma} H_{\sigma\rho\lambda} \,,
\end{equation}
again reproducing the correct kinetic term for the two-form potentials. 

Finally, it is easy to see from \cite{Malek:2016bpu} that the topological term vanishes in the reduction. Thus neither the three-form potential nor its four-form field strength appear in the action as required.

\section{M-theory / Heterotic duality} \label{s:MHetDuality}
Let us now use the results presented here and in \cite{Malek:2016bpu} to discuss the M-theory / heterotic duality in the context of the $\SL{5}$ EFT. Consider the $\SL{5}$ EFT with a K3 surface in the extended space. We can now perform a consistent truncation in two ways.

We let $i, j = 1, \ldots, 4$ label the four coordinates on which the K3 surface depends. These are embedded into $Y^{ab}$ as $Y^{i5}$. We further let $\rho\, \omega_{M,ij}$ be the 22 harmonic two-forms of the K3 surface. Furthermore we take $n^5 = n_5 = 1$ and $\rho = \textrm{constant}$. Thus we have that
\begin{equation}
 \tomega_M{}^{ab} = \rho \epsilon^{abcde} \omega_{M,cd} n_e \,,
\end{equation}
have as their only non-zero components
\begin{equation}
 \tomega_M{}^{ij} = \frac{1}{\sqrt{2}} \rho \epsilon^{ijkl} \omega_{M,kl} \,.
\end{equation}
From the generalised Lie derivative \eqref{eq:LieDerivative} we thus find
\begin{equation}
 \gL_{\tomega_M} \omega_{N,ij} = 0 \,,
\end{equation}
because $\partial_{ij} \tomega_{M,kl} = 0$ and
\begin{equation}
 \gL_{\tomega_M} \omega_{N,i5} = \frac{1}{\sqrt{2}} \epsilon^{klpq} \omega_{N,ik} \partial_l \left( \omega_{M,pq} \rho \right) = 0 \,,
\end{equation}
because $\rho\, \omega_{M,ij}$ are harmonic. We also find that
\begin{equation}
 \gL_{\tomega_M} n^a = 0 \,, \qquad \partial_{ab} n^b = 0 \,,
\end{equation}
and so the doublet and closure condition are satisfied and the embedding tensor and the singlet deformation vanish.

We can choose the four coordinates $Y^{i5}$ as parameterising our section in which case we see that we have performed a consistent truncation of 11-dimensional SUGRA on K3. This way we have obtained an ungauged seven-dimensional SUGRA with $19$ abelian vector multiplets, exactly as required.

However, there is also another interpretation of the above set-up. We could have first performed a reduction of the $\SL{5}$ EFT on K3 and obtained the $\SO{3,19}$ heterotic DFT with abelian gauge group. The DFT fields would then be required to depend only on the six coordinates $Y^{ij}$ since
\begin{equation}
 n^b \partial_{ab} = \left( \partial_{i5}, 0 \right).
\end{equation}
is required to vanish for all reduced fields. Furthermore, we would have obtained 22 twisted derivatives
\begin{equation}
 \partial_M = \frac12 \tomega_M{}^{ab} \partial_{ab} = \frac12 \tomega_M{}^{ij} \partial_{ij} \,.
\end{equation}
Now we could have performed a trivial toroidal truncation of this heterotic DFT, in order to match the previous set-up, where we would have chosen three of the $Y^{ij}$ as parameterising our section. This would have described the consistent truncation of the heterotic string on $T^3$ with the gauge group broken to its abelian subgroup.

From this perspective the difference between the two cases, M-theory on K3 and heterotic on $T^3$ resides in the choice of section: if we take the four coordinates of the K3 surface as our section then we are performing a consistent truncation of 11-dimensional SUGRA on K3, while if we choose one of the other six coordinates as our section then we have more naturally performed a $T^3$ truncation of the abelian $\ON{3,19}$ heterotic SUGRA.

\section{Conclusions}\label{s:Conclusions}
In this paper we have shown that exceptional field theory not only contains 11-dimensional and IIB SUGRA but also the heterotic SUGRA via its doubled version, the heterotic DFT. The EFT can be reduced to the heterotic DFT when its extended space contains a generalised $\SU{2}$-structure manifold. The reduction used is very similar to a truncation on generalised $\SU{2}$-structure manifolds \cite{Malek:2016bpu}. However, the coefficients appearing in the expansion of the EFT fields are still allowed to depend on the extended space, albeit subject to further constraints which ensure that the resulting theory has ${\cal N}=2$ SUSY. The embedding tensor defined by this reduction procedure defines the gauge group of the heterotic DFT while the number of sections of the $\left(\mbf{3},\mbf{1}\right)$-bundle of $\SU{2}_S\times\SU{2}_R \subset \SL{5}$ used in the reduction defines the number of vector fields of the heterotic DFT.

We have shown that using the reduction Ansatz it is natural to introduce $n+3$ ``twisted'' derivatives which commute when the reduced fields are subject to a number of constraints and thus can be thought of as $n+3$ coordinate derivatives. The constraints imposed are also required from the heterotic DFT perspective directly \cite{Hohm:2011ex}. Furthermore, with this Ansatz the generalised Lie derivative of the EFT reduces to that of the heterotic DFT. This suggests that one should interpret the $n+3$ ``twisted'' derivatives as the Fourier duals to momentum and wrapping modes. Indeed, in the case of the consistent truncation of M-theory on K3, we find 22 wrapping derivatives this way.

Finally, we have shown how the duality between M-theory on K3 and the heterotic string on $T^3$ arises in EFT. In the M-theory case, the K3 surface is taken to form the four-dimensional M-theory section of EFT, whereas in the heterotic picture, it is a subset of three out of the extra six directions which are taken as the section. Thus the duality here is generated by a change of section even though there are no isometries. It is thus an example of a ``generalised duality'' without isometries.

It would be interesting to further explore the heterotic / M-theory duality, in particular whether it can capture gauge enhancement when two-cycles of the K3 surface shrink. For example, in \cite{Aldazabal:2015yna} it was shown that double field theory can capture the gauge enhancement at self-dual circles. This would be an interesting test to see whether EFT really captures phenomena which go beyond SUGRA but are related to wrapping sectors of M-theory. Another thing to understand would be whether $\alpha$' corrections are correctly handled in this picture, see for example \cite{Coimbra:2014qaa} for ways these corrections can be incorporated in generalised geometry.

Furthermore, one might ask what happens in lower dimensions. For example, one would expect to see the duality between IIA and the heterotic string in the $\Spin{5,5}$ EFT \cite{Berman:2011pe,Abzalov:2015ega}. In four dimensions one would hope to see mirror symmetry arise between consistent truncations on exceptional $\SU{6}$-structures \cite{Ashmore:2015joa}.

Yet another possible avenue of further research would be to study solutions of exceptional field theory on K3. In \cite{Berman:2014hna}, \cite{Berman:2014jsa} it was shown that the string and pp-wave solutions are unified as one solution of DFT and similarly for branes and monopoles in EFT. How does one describe a solution of EFT on K3 corresponding to an M5-brane wrapping the K3? In this case, its heterotic DFT dual should be the heterotic string and it would be interesting to see how this arises from the formalism presented in this paper.

\section{Acknowledgements}
The author thanks David Berman, Diego Marqu{\'e}s, Carmen Nu{\~n}ez, Felix Rudolph and Henning Samtleben for helpful discussions. The author would also like to thank IAFE Buenos Aires for hospitality while part of this work was completed. This work is supported by the ERC Advanced Grant ``Strings and Gravity" (Grant No. 320045).

\bibliographystyle{JHEP}
\bibliography{NewBib}

\end{document}